# Separate magnetization switching of hexagonal Co/BN/Co junctions grown epitaxially on c-sapphire

**Alexander A.Tonkikh**[1,2,*], **and Peter Werner**[2]

[1]Osram Opto Semiconductors, Leibniz-Str. 4, 93055 Regensburg, Germany
[2]Max Planck Institute of Microstructure Physics, Weinberg 2, 06120 Halle(Saale), Germany

Magnetic tunnel junctions (MTJ) have been grown by using molecular beam epitaxy on c-plane Al2O3 substrates. The MTJ stacks consist of two ferromagnetic hcp-Co layers separated by a thin insulating h-BN barrier. The samples have been grown in a single run revealing single crystalline epitaxial structures with sharp interfaces as observed by applying transmission electron microscopy. The in-plane magnetization experiments have revealed separate magnetization switching of a thin top Co (soft) layer and a thick bottom Co (hard) layer. At zero magnetic field the two Co layers are found in an antiparallel state.

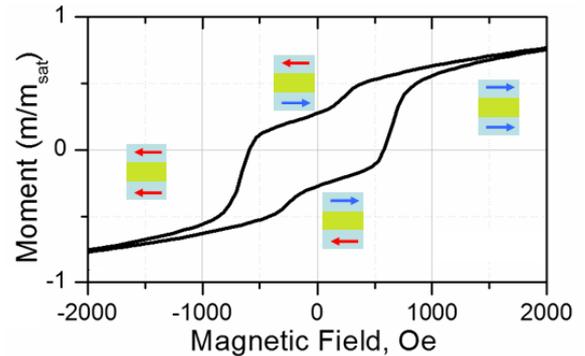

In-plane magnetization switching of an hcp-Co/h-BN-hcp-Co MTJ at room temperature.

**1 Introduction** Magnetic tunnel junctions (MTJs) are important components of the read head of hard disc drives. Moreover, MTJs are considered as promising components of magnetoresistive memory, in particular of the race-track memory [1]. MTJs work on the tunneling magnetoresistance (TMR) principle [2]. They consist of two ferromagnetic layers (hard and soft) separated by a thin insulating barrier. The two ferromagnetic layers could be configured by an external magnetic field or by a current in either parallel or antiparallel configurations. In these two states the resistivity of MTJs is different, being higher in the antiparallel state. Crystalline MgO or amorphous AlOx are used as a dielectric barrier in the state of the art single barrier MTJs providing TMRs from several tens percents for AlOx [3] up to 600% for crystalline MgO [4]. The latter value of 600% became possible due to the spin-filtering effect of crystalline MgO [5]. Multiple advantages of MgO barriers include reliable sputtering fabrication technology, which leads to preferable usage of this barrier type. However, MgO is hygroscopic, which results in the critical reduction of the thermal stability factor ($E/k_bT$) [6]. This deficiency makes scalability of MgO-based MTJs challenging, especially in the sub-40 nm range. An alternative choice for the MTJ barrier could be a crystalline dielectric material, which provides similar spin-filtering properties, however, remains stable at the atmospheric conditions. One of such materials is hexagonal boron nitride (h-BN). Besides its inertness in the atmosphere, h-BN could result in a very high TMR of approximately 1000% in a combination with hcp-Co ferromagnetic electrodes, as calculated for the case of valence band-matched h-BN/hcp-Co [7].

Amorphous BN could also be stable at the atmospheric conditions. However, such barrier reveals only moderate TMR of approximately 25% due to absence of any spin-filtering effect [8]. Relatively high decomposition temperature of BN prevents its re-crystallization inside MTJs in the same way as it works for MgO. Thus, in order to reveal superior spin-filtering and TMR properties, an h-BN-based MTJ should be grown epitaxially. At present there are just a few reports on the epitaxial growth of van der Waals-bonded honeycomb stacks of h-BN monolayers [9-12]. None of them deals with complete MTJ stacks. Therefore, the goal of the present manuscript is to demonstrate the feasibility of the epitaxial fabrication of the h-BN-based MTJ.

One of the technical challenges towards realization of such MTJs is caused by absence of a high-quality substrate, which could be lattice-matched to an h-BN/hcp-Co stack. Single crystal metals, especially Cu(111), could fit due to a close lattice parameter, but appear to be too expensive and rough (rms ~ 10 nm) to become someday a reliable choice for the industry. Therefore, a possible solution for an epitaxial growth of MTJs could be a virtual substrate. Moreover, the virtual substrate should be fabricated on a high quality single crystal basement. One of the possible choices could be sapphire, which has been reported to provide good epitaxial quality of transition metals grown atop [13-15].

---

[*] Corresponding author: email: alexander.a.tonkikh@gmail.com

Another issue is related to the desired type of magnetic anisotropy in MTJs. The state of the art MTJs require perpendicular magnetic anisotropy (PMA), which is necessary for current-induced magnetization switching in the sub-100 nm-range of device lateral dimensions [16, 17]. In the case of our present system (h-BN/hcp-Co), this issue could be fulfilled with Co layers having the thickness of approximately two monolayers. Obviously, such thin Co layers are difficult to grow epitaxially on sapphire due to absence of both, lattice matching and wetting. Slightly thicker Co layers (some nm thick) could be grown on sapphire by depositing Co on a hydroxyl-decorated α-Al2O3 surface [18]. Therefore, the introduction of an extra buffer layer made of a non-ferromagnetic material appears to be more reliable choice. In particular, the PMA of thin Co layers on Ru has been observed and extensively discussed [19]. In our experiments Ru was taken as the intermediate buffer layer material.

In case of our materials of choice, i.e. Co and h-BN, there is a further challenge towards PMA MTJs, which is related to specific epitaxial conditions during metals deposition on the surface of the van der Waals material. At the initial stage, the weekly bonded h-BN structure promotes Volmer-Weber growth mode of Co atop. Such Volmer-Weber growth mechanism has been already reported for metals deposited on a single layer of h-BN [20].

Finally, the above mentioned considerations lead us to the conclusion that MTJs with an in-plane magnetic anisotropy (both Co layers are approximately 10 nm thick) could be a reliable choice for the feasibility test. Moreover, the choice of the Co thickness in the 10 nm-range should guarantee the stable hcp-stacking structure for Co, which could be distorted in case of a 2 monolayers thick Co film [21].

The direct comparison of the c-plane hcp-Ru lattice parameter to the lattice parameter of the c-plane sapphire gives a high mismatch. However, if the Ru lattice is turned around the c-axis by 30 degree, the difference between the double lattice spacing in Ru and the lattice parameter of sapphire reduces down to just 1.5%. Furthermore, the mismatch between hcp-Co and hcp-Ru is approximately 8%, which could be accommodated by strain and lattice defects.

In the present paper, structural and magnetic properties of the MBE-grown samples on Ru-buffered sapphire with a single or double Co layers will be compared. Besides the comparison to a Co/h-BN/Co MTJ stack grown directly on sapphire will be given.

**2 Experiments** The samples were grown by using molecular beam epitaxy (MBE, SIVA 45 setup by Riber) on 1x1 cm$^2$ c-plane (0001) sapphire substrates (CrysTec). A set of three samples will be discussed. Samples 1 and 3 contained two Co layers separated by a thin h-BN layer. Sample 2 contained just a single Co layer. The sequences of layers in these samples were as follows: Sample 1 – Co1-200nm/h-BN-2.7nm/Co2-30nm/Au-10nm; Sample 2 – Ru-15nm/Co1-26nm/h-BN-2.0nm; Sample 3 – Ru-15nm/Co1-26nm/h-BN-2.0nm/Co2-6nm/Ru-3nm. The details of their growth are as follows. Transition metals and gold were deposited in the same MBE growth chamber by using e-beam sources (Riber). Boron was deposited by using a high-temperature effusion cell (Eberl MBE-Komponenten). Nitrogen was supplied by an RF-plasma source (RF-N-50/63, Riber). The source was operated at a background nitrogen pressure in the MBE growth chamber of $1.5 \times 10^{-5}$ torr at a forward RF-power of 250 W. Deposition rates of other materials were as follows: $R_{Co}$= 0.07 Å/s; $R_{Ru}$= 0.015 Å/s; $R_{Au}$= 0.2 Å/s; $R_B$ = 4.4 x $10^{12}$ cm$^{-2}$s$^{-1}$. The latter value corresponds to the h-BN growth rate of approximately 2.8 x $10^{-3}$ ML/s. The deposition rates of metals were preliminary calibrated *in situ* by an oscillating crystal and were verified *ex situ* by using cross-section transmission electron microscopy (TEM). The rate of B was found from a separate experiment on the h-BN/Ni(111) grown structures by using cross-section high-resolution TEM [12].

The substrates were degassed in the MBE chamber prior to the growth at approximately 900 °C for 10 min. The growth of Ru on sapphire has been extensively discussed in the literature. The temperature conditions for Ru/Sapphire were chosen in a similar way as reported in [15]. The growth of the Co1 layer was carried out at the substrate temperature of 500 °C, while h-BN was deposited at 700 °C following the earlier developed procedure [12]. The layers atop h-BN (Co2 and Au or Co2 and Ru) were deposited at room temperature. Ferromagnetic Co1 and Co2 layers were nominally of different thickness in order to ensure their different coercivities. The topmost metal layer, either Au or Ru, was deposited to protect the Co2 layer against oxidation. The latter effect could result in antiferromagnetic cobalt oxide making interpretation of magnetization experiments difficult. During growth, the surface of the samples was monitored by reflection high-energy electron diffraction (RHEED). RHEED patterns were taken in the [11-20] and [1100] directions in respect to the hcp-lattices of Ru or Co. The surface morphology was analyzed by using atomic force microscopy (AFM, Digital Instruments Dimension 5000) in the tapping mode. The microstructure of samples was studied in the cross-section geometry by applying different techniques of transmission electron microscopy: i) high-resolution microscopy (HREM, microscopes TITAN 80/300 and JEM 4010) and ii) scanning transmission microscopy in the HAADF-mode (HAADF-STEM, probe-corrected microscope TITAN 80/300). The TEM specimens were fabricated by applying focused ion beam (FIB) processing. Magnetization experiments were carried out on approximately 2x2 mm$^2$ samples at room temperature in the in-plane geometry in the VSM mode by using SQUID magnetometer (MPMS3 by Quantum Design).

**3 Results and discussion** The samples were *in situ* investigated during MBE growth by using RHEED.

Fig.1 represents RHEED patterns taken at an appropriate stage of the samples growth. All RHEED micrographs demonstrate stripy-patterns, which is characteristic for a flat, epitaxial growth. Moreover, some patterns reveal Kikuchi lines (Fig.1a, b, and d). All RHEED patterns unambiguously indicate single crystalline growth mode of the epitaxial films. It is worth to compare the patterns in Figs.1a (sample 1) and b (sample 3) taken after the Co-1 layer grown a) directly on sapphire (sample 1) and b) on Ru-buffered sapphire. Both patterns are stripy with indication of Kikuchi-lines.

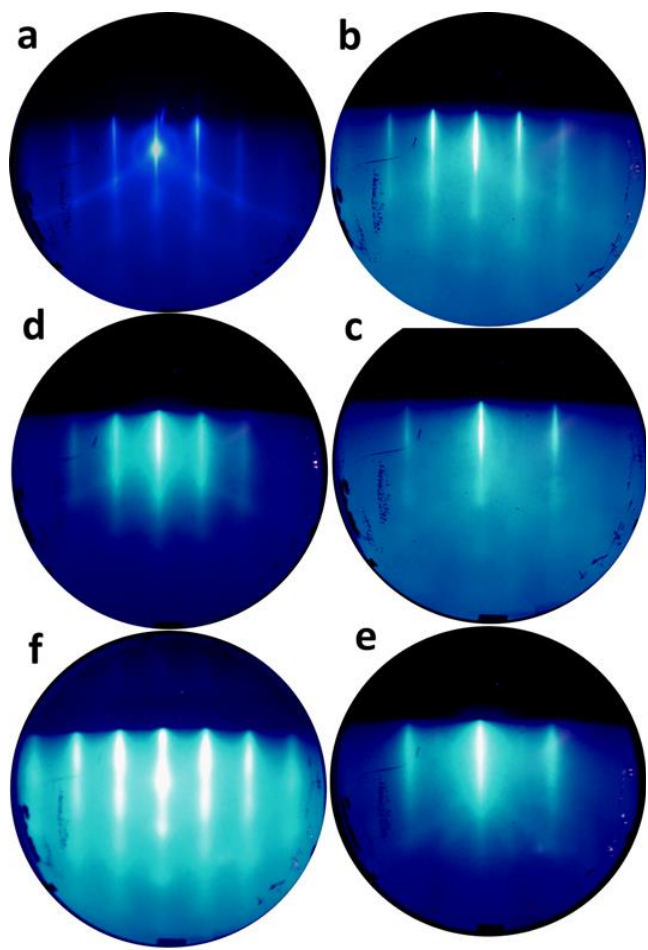

**Figure 1** RHEED patterns taken *in situ* during the growth of samples. The crystallographic directions are related to [11-20] (a, b, d, and f) and [1100] (c and e), respectively ; a) sample 1 after Co-1 layer; b) and c) sample 3 after Co-1 layer; d and e) sample 3 after h-BN layer; f) sample 3 after Co-2 layer.

Therefore, on the basis of this comparison, we can not judge that the structure of the Co-1 layer in sample 3 has a higher crystalline quality, than in sample 1. Figs.1d and 1e show RHEED patterns after the h-BN growth. These patterns nearly completely repeat the patterns of Fig.1b, c (after Co1 layer). This fact demonstrates that the growth of epitaxial h-BN takes place on top of the Co1 layer. The difference in these patterns concerns to thicker reflections after h-BN growth (Figs.1 d, e) in comparison to the hcp-Co surface Fig. 1 b, c). The stripy RHEED pattern in Fig.1f, taken after the deposition of a nominally 6 nm thick Co-2 layer at room temperature, is decorated by spots. Such a pattern could be caused by such a surface, which consists of a single crystalline layer with tiny humps atop. The appearance of such a surface can be explained by an initial Volmer-Weber growth mode of Co on h-BN(0001). The *in situ* RHEED observations have revealed that Co/h-BN islands are generated immediately at the beginning of Co-2 deposition. However, these islands start to merge after approximately 3 nm of deposited Co at the applied growth conditions. After approximately 4 nm of Co-2 deposition, the h-BN surface is completely covered by Co.

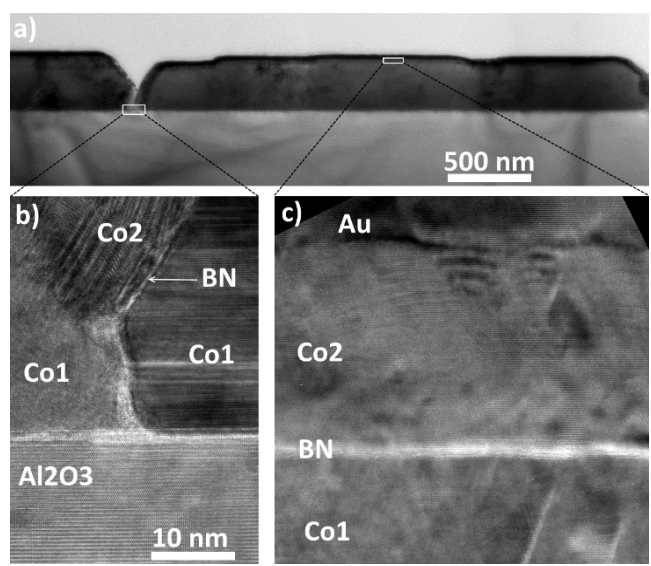

**Figure 2** Cross-section TEM images of sample 1, a) overview image, b) HREM image at the island edge, c) HREM image in the middle of the MTJ-island showing epitaxial structure of Co-1, h-BN, Co-2 and Au layers.

The TEM investigation of sample 1 is depicted in Fig.2. Fig.2a shows in the cross-section that the MTJ stack consists of big flat islands on sapphire. At the applied growth conditions, Co does not wet sapphire that results in an island growth. Furthermore, this growth mode is implemented during the complete stack. It is interesting to compare the structure nearby the h-BN layer at the edge of an MTJ island (Fig2.b) with the structure in the middle of the same island (Fig2.c). At the edge of the island, the plane of the BN layer is not parallel to the sapphire surface indicating that the BN is deposited on the sidewalls of the island. We could not provide evidence that the sidewall-deposited BN layer reveals the hexagonal structure. Moreover, another remarkable phenomenon is revealed at the MTJ island edges: the Co-2 layer of one island could touch the Co-1 layer of a neighboring island forming a metallic contact. Unlike the situation at the MTJ island edges, in the middle of the MTJ-island h-BN grows well separating Co-1 and Co-2 layers (see Fig.2c). It can be assumed that the BN growth conditions on the sidewalls of the Co-island are different from those conditions on the c-plane of hcp-Co. Therefore, we conclude that the direct Co growth on c-sapphire could result in MTJ islands with metallic connections between the bottom and the top ferromagnetic layers. This fact may raise concerns over the possibility of a separate switching of Co1 and Co2 layers. These layers should rather switch in the magnetic field as a single Co layer.

Another situation is observed in case of sample 3, where the MTJ growth takes place on Ru-buffered c-sapphire, as shown in Fig.3. Our results demonstrate that even a 15 nm thick Ru layer could provide a smooth surface suitable for the growth of the MTJ stack. Indeed, the AFM image of sample 3 (Fig.3a) reveals a relatively smooth surface (rms = 1.3 nm) without any presence of MTJ islands. Moreover, the cross-section of this sample, taken in the HAADF-STEM mode (Fig.3b), shows sharp interfaces between all layers of the MTJ stack.

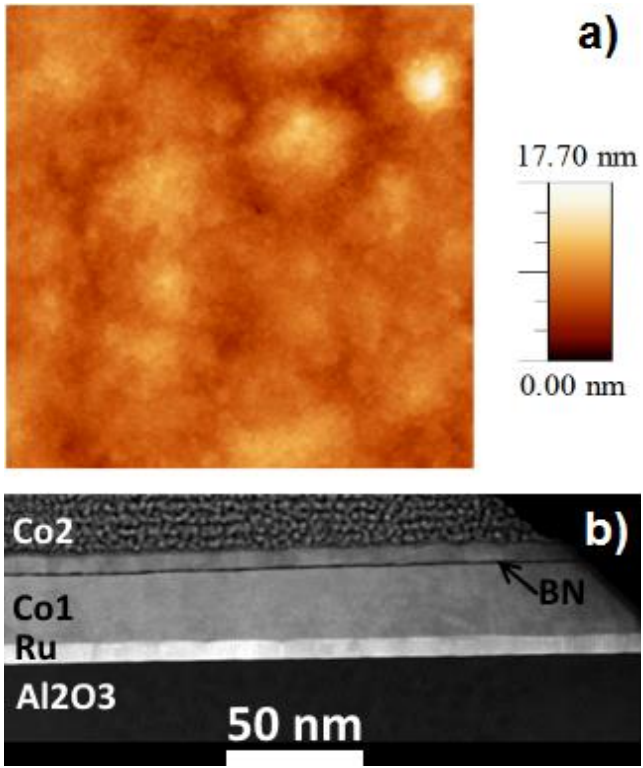

**Figure 3** a) AFM image of sample 3 (area 5 x 5 μm$^2$), b) cross-section HAADF-STEM image of sample 3.

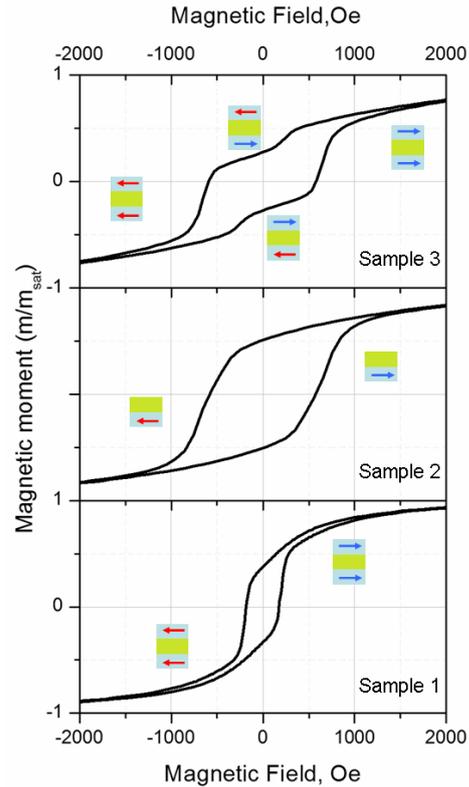

**Figure 4** Magnetization hysteresis loops of samples No. 1 (bottom), No. 2 (middle), and No. 3 (top). The color schemes of magnetization directions in Co-1 and Co-2 layers are given as insets.

The results of the in-plane magnetization experiments are given in Fig.4. Three samples are compared: the reference sample having just a single Co-1 layer (sample 2) and two samples having nominally two Co layers (samples 1 and 3). In case of sample 2 the ferromagnetic hysteresis loop with a well-defined magnetization switching at approximately ±500 Oe is observed. A saturated in-plane magnetization (full curves are not shown here) is achieved at approximately 6 kOe for all three samples. The loop of sample 2 is characterized by a single step magnetization switching, which is typical for the single layer ferromagnetic materials.

A similar single step characteristic is observed in case of sample 1. However, in this case the switching takes place at approximately 200 Oe due to a different thickness of deposited Co. Despite the nominal deposition of two Co layers, separated by an h-BN barrier, no indication of a separate switching is observed. The most feasible reason for that is a direct metallic contact between Co-1 layer and Co-2 layer at the edges of MTJ islands, as observed by cross-section TEM (Fig.2b).

A different situation is observed in case of sample 3. The magnetization curve clearly reveals a double-step switching, which is typical for the MTJs having two ferromagnetic layers [22]. In our case the Co-2 layer is thinner than the Co-1 ones and it could be referred to as a "soft" layer.

Interestingly, this layer switches at a field of B = ±300 Oe, which we interpret as a sign of an antiferromagnetic coupling between Co-1 and Co-2 layers. The bottom Co-1 layer is thick and could be referred to as a "hard" layer. This layer reveals the coercivity of approximately 500 Oe, which is close to the coercivity of Co-1 layer in sample 2 having just one Co-1 layer of the same thickness of 26 nm.

**4 Conclusions** We have demonstrated epitaxial growth and magnetization switching on h-BN based MTJs. The growth of transition metals and a wide band gap semiconductor has been carried out in a single chamber of an MBE system. The separate magnetization switching of hcp-Co/h-BN fully epitaxial MTJs is demonstrated for the first time.

We would like to point out that after an appropriate tuning the direct CIPT measurements along with the conventional magnetoresistance measurements on the e-beam lithography-patterned MTJs could reveal the TMR effect in h-BN based MTJs. The presented results can be regarded as an important step towards a new type of magnetic tunnel junctions for memory applications.

We believe that there are no fundamental difficulties to fabricate similar MTJs by industrial technologies, i.e., chemical vapor deposition (CVD) or magnetron sputtering deposition (MSD), since both these techniques have already been successfully applied for the fabrication of h-BN layers of desired thickness [11, 14].

**Acknowledgements** Valuable discussions with Prof. G.Güntherodt (RWTH Aachen), Prof. S.Parkin (MPI Halle), and Prof. G.Bayreuther (University of Regensburg) are kindly acknowledged. Prof. G.Bayreuther is kindly acknowledged for preliminary VSM-magnetization measurements, H.Blumtritt (MPI Halle) is acknowledged for FIB specimens preparations, D.Bijl (smarttip) is acknowledged for preliminary CIPT-measurements.